\documentclass[utf8]{jetp}
\twocolumn
\usepackage{epstopdf}
\usepackage{xcolor}
\usepackage{graphicx}
\usepackage{cuted} 

\usepackage[colorlinks,
            citecolor=blue,
            anchorcolor=blue,
            menucolor=blue,
            linkcolor=blue,
            filecolor=blue,
            runcolor=blue,
            urlcolor=blue,
            frenchlinks=blue]{hyperref}

\makeatletter

\newenvironment{myreferences}
  {%
    \par\bigskip
    \begin{center}
    REFERENCES
    \end{center}
    \small
    \begin{list}{\arabic{enumiv}.}%
      {%
        \usecounter{enumiv}%
        \setlength{\labelwidth}{1.4em}%
        \setlength{\labelsep}{0.45em}%
        \setlength{\leftmargin}{1.85em}%
        \setlength{\itemindent}{0pt}%
        \setlength{\itemsep}{0pt}%
        \setlength{\parsep}{0pt}%
        \setlength{\topsep}{0pt}%
        \renewcommand{\theenumiv}{\arabic{enumiv}}%
      }%
  }
  {%
    \end{list}%
  }

\makeatother

\begin{document}

\title{Event-by-Event Multiplicity Cumulants as Probes of Partonic Energy Loss and Phase Structure in Au+Au Collisions within a Modified HIJING Model}

\setaffiliation1{Joint Institute for Power and Nuclear Research -- Sosny, National Academy of Sciences of Belarus Acad. Krasina Str. 99, 220109 Minsk, BELARUS}
\setauthor{Y.A.}{Rusak}{1}
\setauthor{L.F.}{Babichev}{}


\abstract{
Event-by-event fluctuations of charged hadron multiplicity, net-proton, and net-electric charge are investigated in central Au+Au collisions at $\sqrt{s_{NN}} =$ 20-200 GeV using a modified HIJING Monte Carlo generator. The model incorporates QCD-inspired partonic energy-loss formalisms in hot (quark--gluon plasma) and cold media, reducing reliance on the default HIJING energy-loss parameter. At energies above 100 GeV/nucleon, multiplicity and net-charge fluctuations show sensitivity to the assumed medium properties, while net-proton results provide an additional comparison with available STAR measurements. By introducing a probabilistic distribution for a first-order phase transition, we find within the present modified-HIJING framework that multiplicity cumulant ratios ($\sigma^2/M$ and $S\sigma$) show a pronounced non-linear response to the adopted phase-allocation ansatz
near the selected binodal energy, whereas net-charge and net-proton fluctuations remain relatively insensitive. Furthermore, expanding the kinematic acceptance enhances the sensitivity of multiplicity fluctuations to the selected transition region in the present phenomenological ansatz. 
}

\maketitle

\section{Introduction}
Searching for quark-gluon plasma (QGP) has been one of the principal goals of relativistic heavy-ion experiments over the last decades. At vanishing baryon chemical potential, lattice QCD calculations with physical quark masses establish that the transition from hadronic matter to QGP is an analytic crossover~\cite{Aoki06}. At finite net-baryon density, a direct first-principles determination of the QCD phase structure remains difficult because of the sign problem. A first-order transition line and a critical endpoint (CEP) at sufficiently large $\mu_B$ are possible scenarios suggested by effective models and phenomenological studies, but their location and even their existence remain open questions~\cite{Sun18}.

The collision energy interval considered in this work includes conditions for which the QCD transition is expected to be a crossover at sufficiently small baryon chemical potential. Therefore, the first-order-transition treatment introduced below is not intended as a claim that a physical first-order transition occurs throughout the RHIC energy range. Instead, it is used as a phenomenological finite-volume sensitivity ansatz to examine how fluctuation observables respond to a smooth event-by-event interpolation between cold and hot generator scenarios.

Therefore, observables are needed that are both sensitive to the properties of the created medium and experimentally accessible ~\cite{Blume18,Biswarup17,Shuryak17,Koch17}. Heavy-ion measurements at the SPS (NA49), RHIC (PHENIX and STAR), and the LHC (ALICE and CMS) have investigated particle production, correlations, and event-by-event fluctuations over a broad range of collision energies ~\cite{Adare08,Gazdzicki08,Xu14,Luo15,ALICE21,CMS23}. Future measurements at FAIR, in particular with the CBM experiment, and at the NICA collider with the MPD experiment will extend these studies
towards lower collision energies and larger baryon densities
\cite{Kekelidze18,FAIRNICA}.

First-order phase transitions are associated with mixed-phase formation and discontinuous thermodynamic quantities in the thermodynamic limit. In finite and rapidly evolving heavy-ion systems, the corresponding signals can be smeared by finite-size and dynamical effects. Event-by-event and local fluctuations in particle production have therefore  been proposed as possible probes of critical and first-order-transition phenomena ~\cite{Mohanty02,Yang98,Hippert17}. \par 

To predict experimental results on phase transition observables, Monte-Carlo simulations are used. There are generators that are based on 3 types of physical models: transport, hydrodynamical, and hybrid. Transport models are phenomenological and possible phase transitions can be implemented there by either changing values of some parameters or turning on/off physical processes (jet quenching, different state radiations, etc.). Hydrodynamical models use different equations of state, where the phase transition can be implemented explicitly with a change of thermodynamical quantities. Hybrid models combine hydrodynamic evolution with a microscopic hadronic afterburner, allowing a realistic description of both the bulk and the late‑stage hadronic scattering (e.g. UrQMD~\cite{Bleicher99} or JAM~\cite{Nara19} as afterburners).

A characteristic hard-probe signature of a hot and dense QCD medium is partonic energy loss during propagation through the medium
~\cite{Armesto12}. Thus for modeling of QGP produced in heavy-ion collisions we modified the HIJING (Heavy Ion Jet INteraction Generator) v.1.411~\cite{Gyulassy94} transport model according to parton energy loss formalisms in both finite- and infinite-sized media~\cite{Baier95, Baier97}. In this way, we reduce reliance on the default HIJING energy-loss parameter by replacing it with QCD-inspired expressions linked to the assumed medium properties. In addition, we implemented the parton energy loss model in a cold medium ~\cite{Mueller96} for comparison. After that we analyzed event-by-event fluctuations and the possibility for them to identify different types of matter and phase transitions in heavy-ion collisions in different kinematic regions. We compared results on fluctuations for hot and cold media, default HIJING with the default energy loss $\frac{dE}{dz} =$ 2 GeV/fm and for the case without quenching. We also study the behavior of event-by-event fluctuations within a phenomenological first-order-transition ansatz that interpolates between cold-matter and hot-medium generator scenarios. Within the present modified-HIJING framework, charged-hadron multiplicity fluctuations provide greater sensitivity than the considered net-charge and net-proton observables to the assumed energy-loss scenarios and to the adopted phase-allocation ansatz.

\section{Generator modification}
HIJING is a PYTHIA-based Monte Carlo generator. It combines a QCD-inspired model for jet production with the Lund model for jet fragmentation. The formulation of HIJING was guided by the Lund FRITIOF and Dual Parton models for soft $A + B$ reactions at intermediate energies ($\sqrt{s} <\sim 20$~GeV/nucleon) and the successful implementation of pQCD processes in the PYTHIA model for hadronic collisions. HIJING is designed mainly to explore the range of possible initial conditions that may occur in relativistic heavy-ion collisions. To study nuclear effects, it also includes nuclear shadowing of parton structure functions and a schematic model of the final state interaction of high PT jets in terms of an effective energy loss parameter, $dE/dz$. At $pp$ and $p\bar p$  levels, HIJING also made an important effort to address the interplay between low $p_T$ nonperturbative physics and  hard pQCD processes. This Monte Carlo model has been extensively tested  against data on $p + p(\bar p)$ over a wide energy range, $\sqrt{s} = 50-1800$~GeV, and $p + A, ~A + A$ collisions at moderate energies $\sqrt{s} \leq 20$~GeV/n. 

From the phenomenological point of view, a viscous hydrodynamics description combined with a hadronic afterburner provides the most realistic framework for bulk soft observables and azimuthal anisotropies at RHIC and LHC energies. Nevertheless, transport and jet-based generators such as HIJING and UrQMD remain widely used as baselines for fluctuation studies and for exploring phenomenological implementations of energy-loss and phase-transition scenarios over broad ranges of $\sqrt{s_{NN}}$~\cite{Armesto12,Stachel07}. In the present work, we therefore use a modified HIJING as a flexible transport framework in which QCD-inspired partonic energy-loss and a finite-volume phase transition ansatz can be implemented event-by-event, rather than attempting a detailed hydrodynamic description of all soft observables. The present modification of the parton energy-loss model is described below.

Heavy-ion collisions can produce hot, dense matter (i.e. QGP) and a cold hadronic medium. The energy loss of a parton in a hot medium consists of two parts: energy loss due to radiation and energy loss due to collisions of the parton with particles of the medium.

\begin{equation}\label{eq:1}
    \left.\frac{dE}{dz}\right|^{total}=\left.\frac{dE}{dz}\right|^{collisional}+\left.\frac{dE}{dz}\right|^{radiative}\,.
\end{equation}

Radiation energy losses have the form~\cite{Baier95,Baier97}:
\begin{wide}
\begin{equation}\label{eq:2}
\hspace*{-10pt}\frac{dE}{dz}\bigg|^{radiative}=
\begin{cases}
 \frac{\alpha_s C_R}{8}\frac{\mu^2}{\lambda_{g}}L\ln{\frac{L}{\lambda_g}} & \mbox{$(L<L_{CR})$}~(BDMPS - approach)
\\
\\
\frac{\alpha_s C_R}{8}\sqrt{\frac{E\mu^2}{\lambda_g}}\ln{\frac{E}{\lambda_g \mu^2}} & \mbox{$(L>L_{CR})$}
\end{cases}\hspace*{-3pt}\,,
\end{equation}

In turn, part of the losses due to collisions~\cite{Piegne08}:

\begin{equation}\label{eq:3}
\hspace*{-10pt}\frac{dE}{dz}\bigg|^{collisional}\hspace*{-8pt}=
\hspace*{-3pt}\begin{cases} \displaystyle C_R\pi\alpha_s^2
T^2 \left[ \left( A_g+A_q\frac{N^{eff}_f}{6} \right) \ln{\frac{ET}{m_D^2}}+O(1) \right] - \hbox{u,d,s,g partons}
\\
\\
\frac{4\pi\alpha^2_sT^2}{3}\left[\left(A_g+A_q\frac{N^{eff}_f}{6}\right)\ln{\frac{ET}{m_D^2}}+\displaystyle \frac{2}{9}\ln{\frac{ET}{M^2}}+c(N^{eff}_f)\right]
- \hbox{c,b,t partons}
\end{cases}\,,
\end{equation}

\end{wide}

\noindent
where $\mu^2=4\pi\alpha_sT^2(A_g+A_qN^{eff}_f/6)$  is the  squared Debye screening mass of the medium, $C_R=C_g=N^{eff}_c$  is the gluon color factor, $C_R=C_q=((N^{eff}_c)^2-1)/(2N^{eff}_c)$ is the quark color factor, $A_g$ and $A_q$ are the quantities characterizing the chemical equilibrium of the system (we take 1 to mean that the system is in equilibrium),   $\lambda_g=1/(4\pi\alpha_sT)$ is the  mean free path of the gluon and $L=3r_0A^{1/3}/4, ~r_0=1.112~fm$  is the transverse length of the resulting medium. In addition, the energy loss due to radiation is characterized by $L_{CR}=\sqrt{\frac{\lambda_gE}{\mu^2}}$  , which is a quantity that characterizes the finiteness/infinity of the medium for the flying parton. In turn, the effective number of flavors, the effective number of colors, and the strong interaction constant, which depends on the temperature, are expressed as~\cite{Steffens06}:

\begin{equation*}\label{eq:4}
 N_f^{eff}=\left[1-2\pi^2\frac{T^2/Q^2-T^2/Q_0^2}{\ln{\left(\frac{Q^2}{Q^2_0}\right)}}\right]N_f\,,
\end{equation*}

\begin{equation*}\label{eq:5}
 N_c^{eff}=\left[1+\frac{8\pi^2}{11}\frac{T^2/Q^2-T^2/Q_0^2}{\ln{\left(\frac{Q^2}{Q^2_0}\right)}}\right]N_c\,,
\end{equation*}

\begin{equation*}\label{eq:6}
\begin{split}
    &  \alpha_s= \alpha_s(Q,T)= \\[3pt]
   = & \frac{\alpha_s(Q_0)}{1+\frac{\alpha_s(Q_0)}{4\pi}\left[\frac{11}{3}N_c^{eff}-\frac{2}{3}N_f^{eff}\right]\ln{\left(\frac{Q^2}{Q_0^2}\right)}}\, ,\\[8pt]
    & \quad N_c=3, \quad Q_0=m_Z=91~\hbox{GeV}\,.
\end{split}
\end{equation*}

The dependence of the temperature of the medium on the collision energy was obtained by fitting data from hydrodynamic calculations for various experiments~\cite{Stachel07} and is expressed as  \mbox{$T_0=0.124\left(\sqrt{s_{NN}}\right)^{0.222}$~GeV} with the temperature profile of the medium \mbox{$T=T_0\left[2\left(1-\frac{r^2_\bot}{L^2}\right)\right]^{1/4}$.} For a cold medium, we use only  energy loss of the parton due to radiation, the expression for which and all parameters are presented, for example, in~\cite{Mueller96}.

In the modified HIJING scheme, the radiative energy loss
$\Delta E_{\rm rad}$ and collisional energy loss $\Delta E_{\rm coll}$ are computed for each high-$p_T$ parton along its path in the medium. The resulting reduced parton energy is passed to the standard Lund fragmentation routine. Therefore, the present implementation should be regarded as a phenomenological treatment of the coupling between partonic energy loss and final-state multiplicity, rather than a complete microscopic description of medium response.

\begin{figure}[h!]
\leavevmode
    \centering
     \includegraphics[angle=0, width = 1.\linewidth]{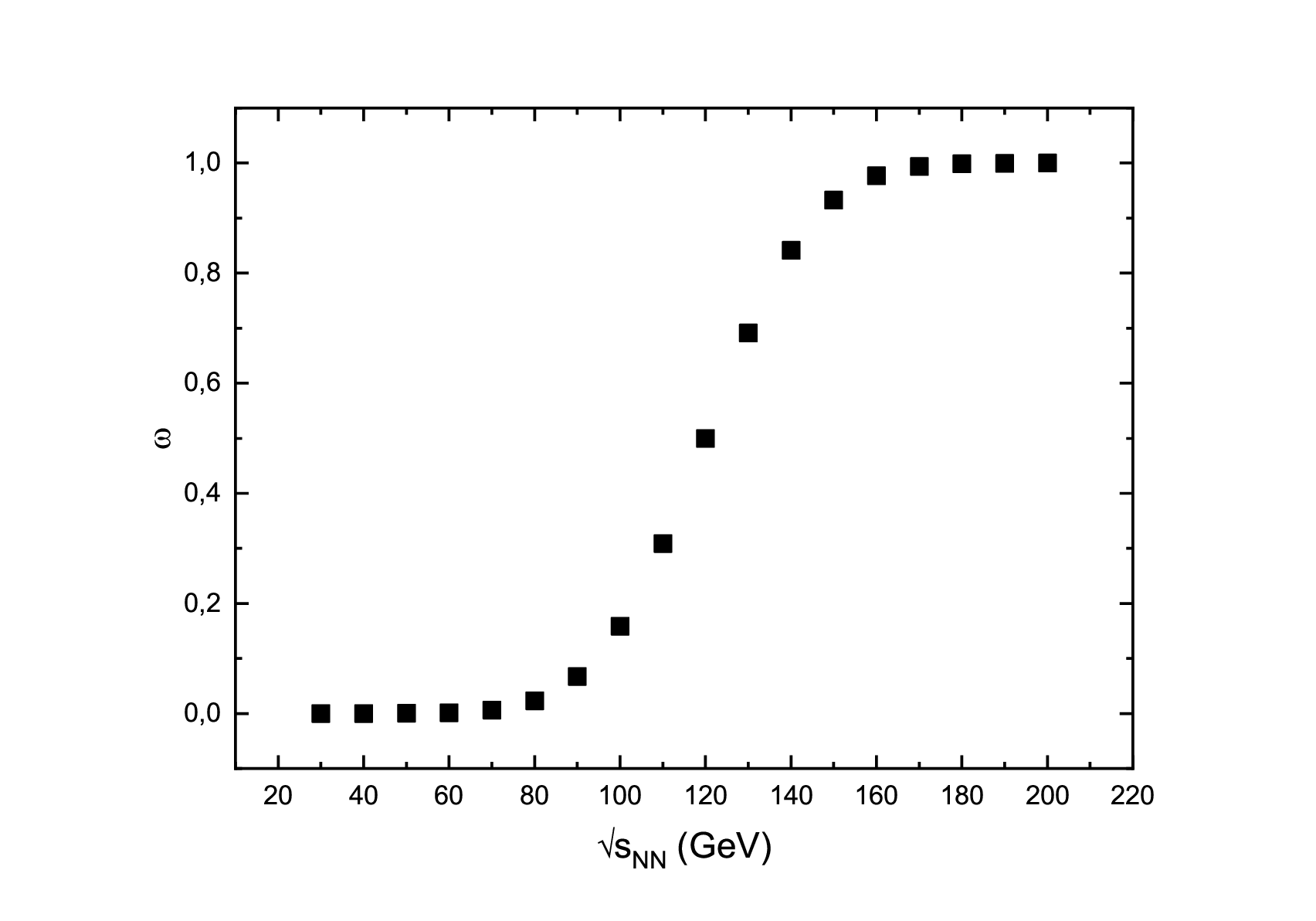}
      \caption{(Color online) Example of the probability distribution of first order phase transition in event-by-event collisions versus $\sqrt{s_{NN}}$. The parameters of the distribution are: $(\sqrt{s_{NN}})_{b}=120$~GeV, $\sigma=20$~ GeV.}
     \label{fig:1}
  \end{figure}

To incorporate a finite-volume first-order transition into the generator, we introduce a phenomenological phase-allocation probability $\omega(\sqrt{s_{NN}})$ of error-function form ~\cite{Rusak20} (example of such distribution is on the Figure \ref{fig:1}):
\begin{equation}\label{eq:7}
\begin{split}
   & \omega_i(x_i)=\frac{1}{2}\left[1+\hbox{erf}\left(\frac{x_i}{\sqrt{2}\sigma}\right)\right] \,,\\[8pt]
    & x_i=\left(\sqrt{s_{NN}}\right)_i -
    \left(\sqrt{s_{NN}}\right)_b\,,
\end{split}
\end{equation}
\noindent

This choice provides a smooth interpolation between hadronic and deconfined regimes in finite, rapidly expanding heavy-ion fireballs, where the sharp discontinuity expected in the thermodynamic limit is rounded by finite-size and dynamical effects. In this implementation, $\sqrt{s_{NN}}$ acts as the external control parameter, $\sqrt{s_{NN}}_b$ denotes the binodal energy, and $\sigma$ is an effective width that characterizes the smearing of the transition region. The error-function dependence is adopted here as a phenomenological ansatz at the generator level, rather than as a direct consequence of the cited thermodynamic arguments. This construction is consistent with the general picture of finite-size effects and spinodal decomposition in nuclear systems~\cite{Chomaz04, Randrup09, Randrup10}, while the fluctuation framework of Landau and Lifshitz~\cite{Landau76} provides the broader statistical context for understanding fluctuations and finite-size rounding in phase transitions. \par
We implement $\omega(\sqrt{s_{NN}})$ event-by-event: for each generated collision at $\sqrt{s_{NN}}$ we draw a uniform random number $r\in[0,1)$ and assign the event to the hot matter branch if $r<\omega(\sqrt{s_{NN}})$ and to the cold matter branch otherwise. This probabilistic allocation preserves generator modularity and allows straightforward sensitivity studies of $\sigma$ and $\sqrt{s_{NN}}_b$ on fluctuation observables. The two branches differ only in the energy loss mechanisms described above, while the remaining HIJING settings are kept at default values.

\section{Results and discussion}
This section presents the results of the analysis of the multiplicity fluctuations of the charged hadron ($K^{+}, K^{-}, \pi^{+}, \pi^{-}, p, \bar{p}$) from simulations of central Au+Au collisions in the energy range of $\sqrt{s_{NN}}$ $=$ 20-200 GeV. Central events are selected by restricting the impact parameter to the range $b=0-3.4$ fm, which corresponds to approximately the $0-5\%$ most central collisions in a Glauber-type geometric picture for Au+Au at these energies. The number of events is 6.5 million for each data point in the figures. 

We apply the pQCD‑based energy‑loss formalism in a RHIC energy range where $\sqrt{s_{NN}}=20-200$ GeV and typical hard scattering scales $Q^{2}$ are several GeV$^2$. This is admittedly at the lower end of strict pQCD applicability, and our approach should be viewed as a phenomenological implementation guided by QCD‑inspired energy‑loss models rather than a precise perturbative calculation.

In what follows, we consider event-by-event fluctuations of a generic observable $k$ (net-proton number $P=N_p-N_{\bar p}$, net electric charge $Q=N_+-N_-$, and charged-hadron multiplicity $N=N_++N_-$), where $N_+$ and $N_-$ denote the numbers of accepted positively and negatively charged final- state particles, respectively. For each distribution we define:
\begin{equation}
\begin{split}
& C_1^k = \langle k \rangle = M_k \,, \\
& C_2^k = \langle (\delta k)^2 \rangle = \sigma_k^2 \,, \\
& C_3^k = \langle (\delta k)^3 \rangle \,, \\
& S_k = \frac{C_3^k}{(C_2^k)^{3/2}} \,,
\end{split}
\end{equation}
where \(\delta k = k_i - \langle k \rangle\) for the \(i\)-th event, and \(\langle \cdots \rangle\) denotes the average over all events in the sample. The cumulant ratios used in the figures are

\begin{equation}
\frac{C_2^k}{C_1^k}=\frac{\sigma_k^2}{M_k}\,,
\qquad
\frac{C_3^k}{C_2^k}=S_k\sigma_k\,.
\end{equation}

Statistical uncertainties of the cumulant ratios are estimated using large-sample delta-theorem expressions~\cite{Zhang22,Chen17}. For each $\sqrt{s_{NN}}$ and kinematic window, $N=6.5\times10^6$ events are generated. The uncertainty of $C_2^k/C_1^k$ is evaluated from the variances of the sample mean and variance. For  $C_3^k/C_2^k$, we use a Gaussian-limit estimate as a statistical baseline. Because the simulated distributions can be non-Gaussian, a full uncertainty determination would require higher central moments or bootstrap resampling. The corresponding relative statistical uncertainties are stated in the figure captions.

\begin{figure}[tpb]
\leavevmode
    \centering
     \includegraphics[angle=0, width = 1.0\linewidth]{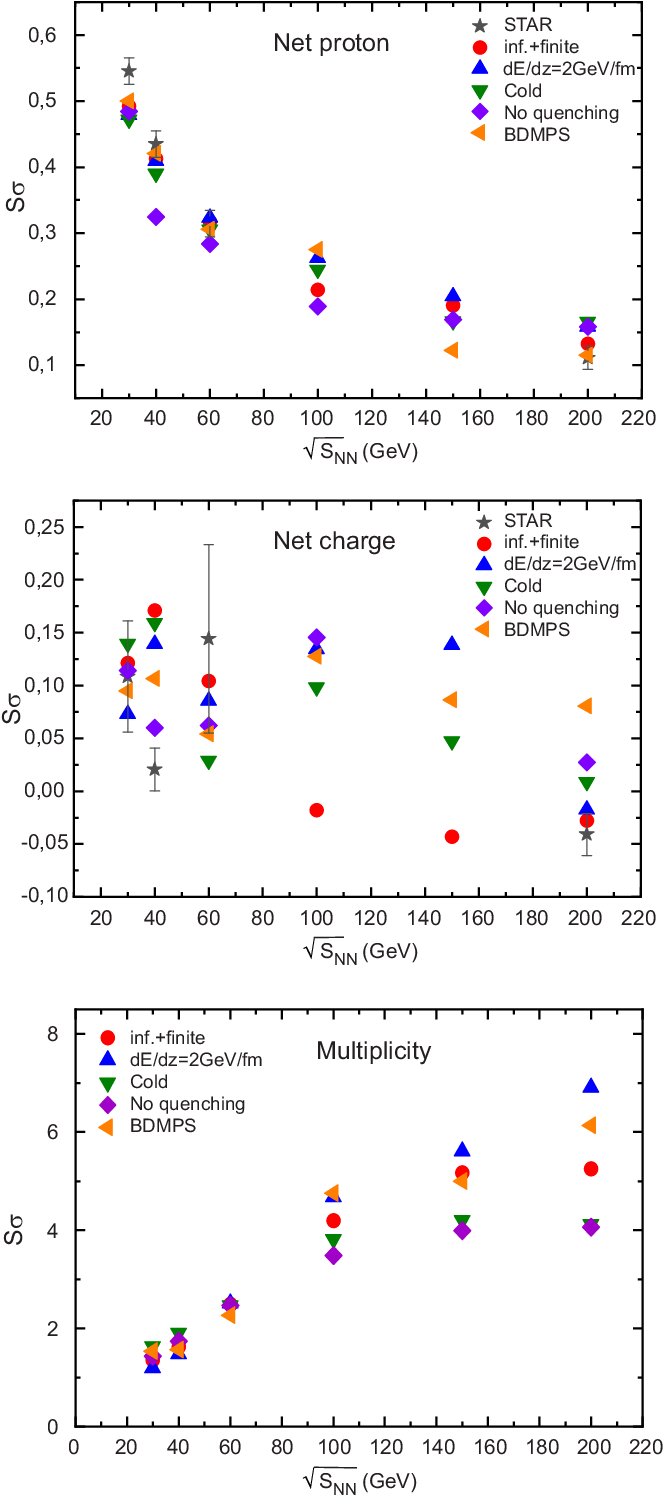}
      \caption{(Color online) The ratio of the 3rd to 2nd cumulant of net-proton, net-electric and charged hadron multiplicity fluctuations vs $\sqrt{s_{NN}}$ for different models of parton energy loss in medium: inf.+finite - full radiative-plus-collisional energy-loss model according to Eq.~\ref{eq:2} and Eq.~\ref{eq:3}; cold - cold matter~\cite{Mueller96}; BDMPS - BDMPS model only; 2 GeV/fm - unmodified HIJING with a parton energy loss parameter of 2 (default) ; no quenching - no medium formation effects, STAR - STAR experiment results. The kinematic cuts are: $|y|<0.5$; $0.2<p_{T}<2.0$ GeV/c. The largest estimated relative statistical uncertainties among the displayed model points are approximately 4\% for net-proton, 9.5\% for net charge, and 1.2\% for charged hadron multiplicity.}
     \label{fig:2}
  \end{figure}

Figure \ref{fig:2} shows the energy dependence of the ratio of the third cumulant to the second cumulant of net-proton, net-electric and charged hadron multiplicity. The simulations for each energy were made with different medium parameters which are expressed by abbreviations that are used on the graphs: inf.+finite – full radiative-plus-collisional energy-loss model according to
Eq.~\ref{eq:2} and Eq.~\ref{eq:3}; cold - cold matter~\cite{Mueller96}; BDMPS - BDMPS model only; 2 GeV/fm - unmodified HIJING with a parton energy loss parameter of 2 (default) ; no quenching - no medium formation effects. The kinematic cuts are the following: $|y|<0.5$; $0.2<p_{T}<2.0$ GeV/c that correspond to the kinematic cuts of the STAR experiment~\cite{STAR25}. The STAR data points were taken from~\cite{Luo15}. 
\wideon
  \wideoff

The figure demonstrates that at lower energies parton energy loss mechanism doesn't dominate, but at ultrarelativistic energies ($>$100 GeV/nucleon) parton energy loss across all considered models becomes sufficiently pronounced for fluctuations to serve as a viable probe of the properties of the medium produced in heavy-ion collisions. At the highest RHIC energies, the net-charge and net-proton results are qualitatively consistent with the hot-medium scenarios, particularly those including finite-size effects. In contrast to net-charge fluctuations, which decrease with increasing collision energy in the present calculations, multiplicity fluctuations increase with the assumed medium temperature. Within this model, they are therefore useful for discriminating among the considered energy-loss scenarios.\par

\begin{figure}[htbp]
\leavevmode
    \centering
     \includegraphics[angle=0, width = 1.\linewidth]{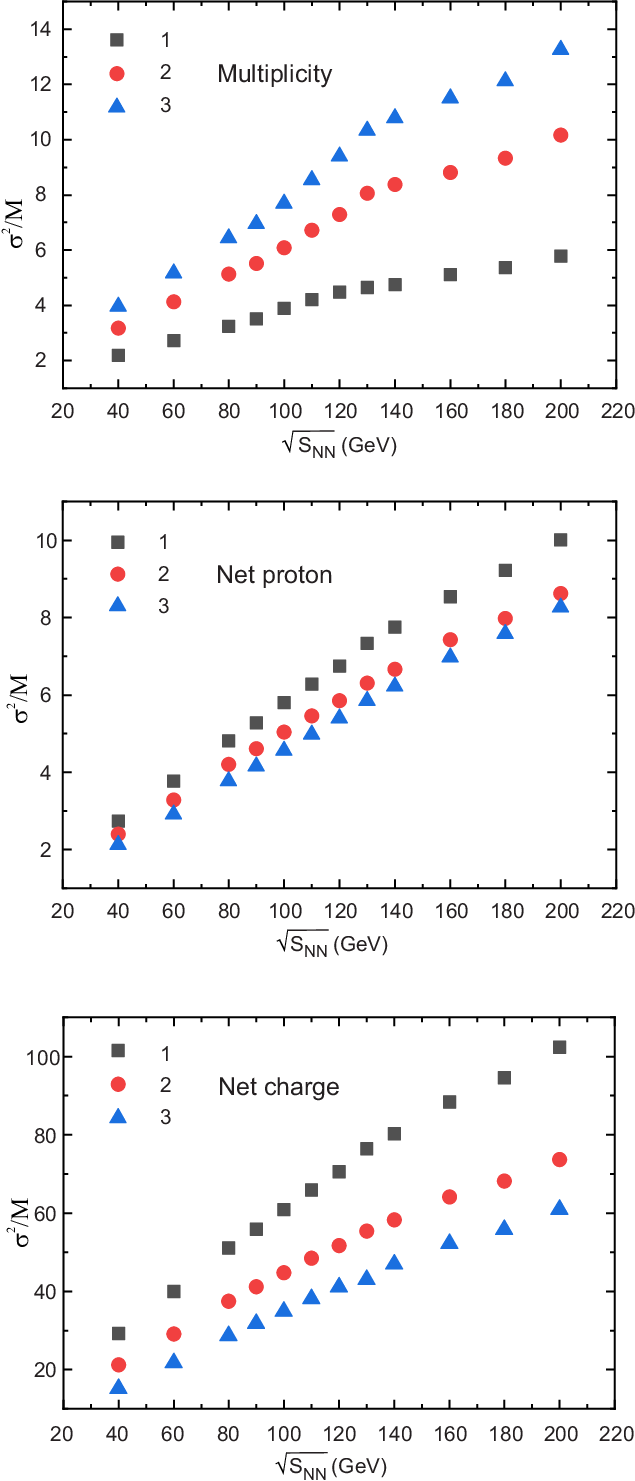}
      \caption{(Color online) The ratio of the 2nd to 1st cumulant of charged hadron multiplicity, net-charge and net-proton number vs $\sqrt{s_{NN}}$ in different kinematic regions (1 -- $|y|<0.5$, $0.2<p_{T}<2.0$ GeV/c; 2 --  $|y|<1.0$; $0.2<p_{T}<2.0$ GeV/c; 3 -- $|y|<1.0$; $0<p_{T}<2.0$ GeV/c.) using the phenomenological phase-allocation probability for the first-order-transition ansatz.  The parameters of this distribution are: $(\sqrt{s_{NN}})_{b}=120$~GeV, $\sigma=20$~ GeV. The largest estimated relative statistical uncertainties  among the displayed model points are approximately 0.07\% for net-proton, 0.12\% for net charge, and 0.06\% for charged hadron multiplicity.}
     \label{fig:3}
  \end{figure}

Figure \ref{fig:3} shows the dependence of the ratio of the second to first charged hadron cumulants on the probability distribution (Fig.~\ref{fig:1}). The parameters of this distribution are: $(\sqrt{s_{NN}})_{b}=120$~GeV, $\sigma=20$~ GeV. This dependence was analyzed in 3 different kinematic regions:
\begin{enumerate}
\item $|y|<0.5$; $0.2<p_{T}<2.0$ GeV/c;
\item $|y|<1.0$; $0.2<p_{T}<2.0$ GeV/c;
\item $|y|<1.0$; $0<p_{T}<2.0$ GeV/c.
\end{enumerate}

Figure~4 presents the $S\sigma=C_3/C_2$ ratios for net-proton and multiplicity fluctuations in the same three kinematic regions.

  \begin{figure}[htbp]
\leavevmode
    \centering
     \includegraphics[angle=0, width = 1.0\linewidth]{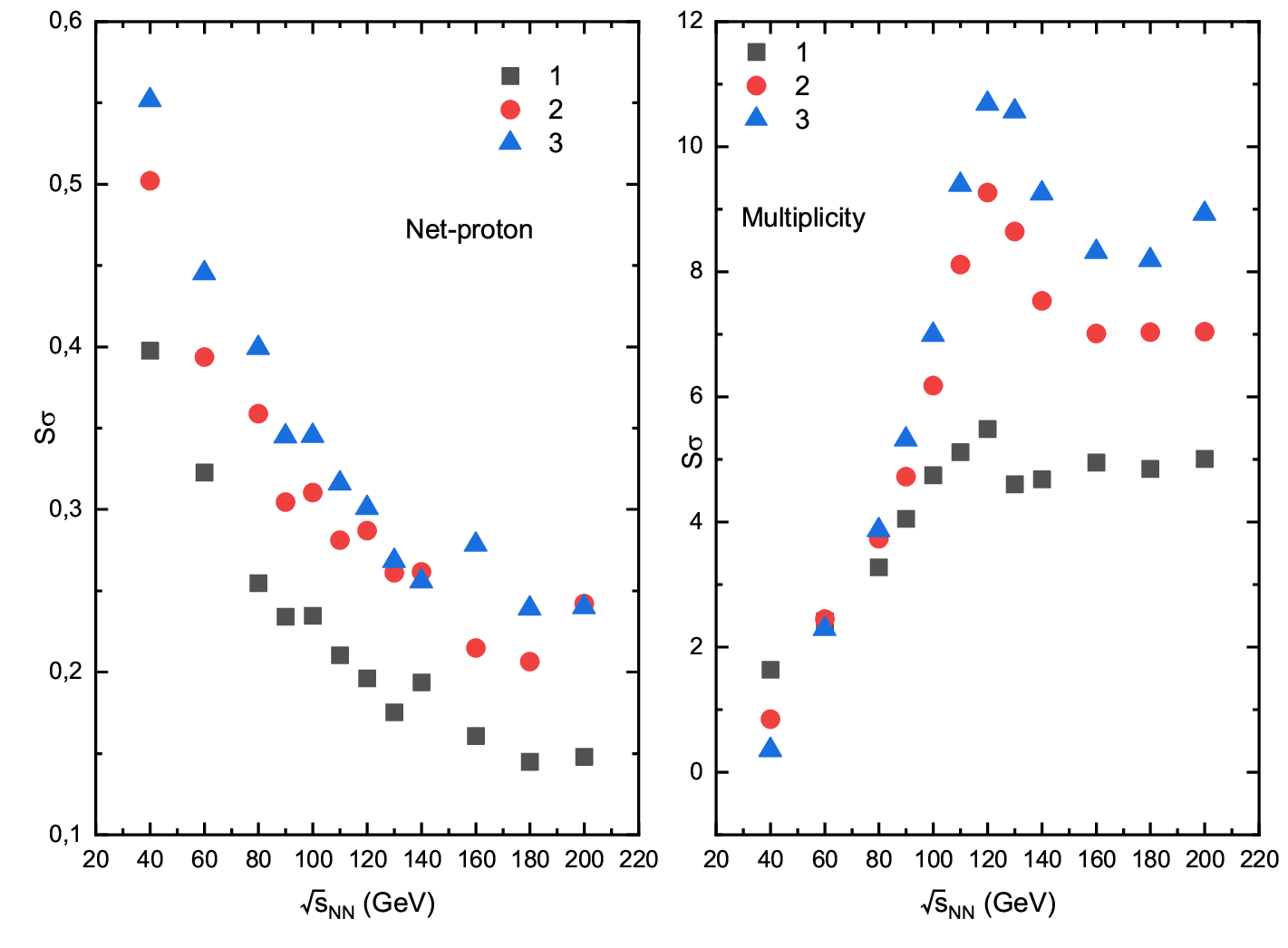}
      \caption{(Color online) The ratio of the 3rd to 2nd cumulant of charged hadron multiplicity and net-proton number vs $\sqrt{s_{NN}}$ in different kinematic regions (1 -- $|y|<0.5$, $0.2<p_{T}<2.0$ GeV/c; 2 --  $|y|<1.0$; $0.2<p_{T}<2.0$ GeV/c; 3 -- $|y|<1.0$; $0<p_{T}<2.0$ GeV/c.) using the phenomenological phase-allocation probability for the first-order-transition ansatz.  The parameters of this distribution are: $(\sqrt{s_{NN}})_{b}=120$~GeV, $\sigma=20$~ GeV. The largest estimated relative statistical uncertainties are approximately
      $3\%$ for net-proton  and $1.7\%$ for charged hadron multiplicity.}
     \label{fig:4}
  \end{figure}
The results demonstrate that charged hadron multiplicity fluctuations are more sensitive to the implemented phase-allocation ansatz than the considered net-charge and net-proton observables, regardless of the cumulant order. However, the response and accompanying non-linear behavior are most pronounced in the third-to-second cumulant ratio (\(S\sigma\)). Furthermore, expanding the kinematic acceptance markedly enhances the sensitivity of these multiplicity fluctuations near the selected transition region of the phase-allocation ansatz. In contrast, net-proton cumulants exhibit divergent trends with broadening kinematic coverage, where the \(\sigma^2/M\) ratio decreases while \(S\sigma\) increases.\par

The stronger response of charged-hadron multiplicity follows from its inclusive character: changes in the energy-loss prescription modify the production of soft charged particles and therefore affect the total accepted multiplicity event by event. In contrast, net-charge and net-proton observables involve cancellations between positive and negative particles or between protons and antiprotons. Their sensitivity to the branch allocation is consequently reduced by these cancellations and by conservation constraints.

Part of the weaker response observed for net-proton and net-charge cumulants in Figures 2-4 can be attributed to acceptance and conservation constraints.\par

Published charged-hadron multiplicity-fluctuation measurements from PHENIX~\cite{Adare08} are not directly overlaid on the present results. That analysis reported the scaled variance $\omega_{\rm ch}=C_2/C_1$ for Au+Au
collisions at $\sqrt{s_{NN}}=62.4$ and 200 GeV, whereas Fig.~2 presents the ratio $C_3/C_2$. In addition, the PHENIX measurement used the central-arm acceptance $|\eta|<0.26$ with a limited azimuthal coverage and included detector-efficiency and geometry-fluctuation corrections.
These conditions differ from the present generator-level calculation with $|y|<0.5$ and the stated transverse-momentum cuts; therefore, a quantitative comparison is not attempted.

The modified hot-medium scenario differs from the default and cold-matter cases because the partonic energy loss is linked to the assumed local medium properties. Radiative and collisional losses depend on the local temperature, path length, running $\alpha_s$, collision geometry, and $\sqrt{s_{NN}}$. In the present implementation, the resulting reduced parton energy is passed to Lund fragmentation. Therefore, the observed changes of multiplicity cumulants should be interpreted as a phenomenological response of the modified HIJING model to the assumed energy-loss prescription, rather than as a complete microscopic description of medium response.

We note that, in a realistic experimental environment, charged-particle detection cannot reach arbitrarily small transverse momenta: very low-\(p_T\) particles can spiral entirely within the beam pipe due to the magnetic field and remain undetected. Our kinematic region with $0 < p_T < 2.0$~GeV/c should therefore be regarded as an idealized acceptance illustrating the sensitivity of multiplicity cumulants to enlarged phase-space coverage, while realistic lower $p_T$ cuts, for example $p_T>0.2$~GeV/$c$, will somewhat reduce this sensitivity. No detector response or tracking-efficiency correction is applied in the present generator-level study; the kinematic cuts are used only to define the accepted final-state particles.

The present analysis is restricted to approximately $0-5\%$ central Au+Au collisions in order to reduce event-by-event variations of the collision geometry and participant number. In noncentral collisions, additional volume, path-length, and participant fluctuations are expected to affect
the cumulants and may obscure the sensitivity to the implemented energy-loss and phase-allocation prescriptions. A systematic extension to noncentral events would therefore require a dedicated treatment of centrality-bin-width effects and geometry fluctuations
\cite{Luo2015CBWC,STAR2016Cumulants,Adare08}.

An extension of the modified-HIJING framework to LHC energies is possible in principle. However, such an extension would require a separate validation and retuning of the underlying HIJING generator parameters, which are used here in a configuration intended for the RHIC energy range. 
In particular, the parameters controlling hard-process production and its minimum transverse-momentum scale, nuclear parton shadowing, initial parton transverse momentum, string fragmentation, and the medium energy-loss prescription would require dedicated validation at LHC energies.

\section{Conclusions}
In summary, this work investigates event-by-event multiplicity and conserved charge fluctuations in central \(\text{Au}+\text{Au}\) collisions at center-of-mass energies of $\sqrt{s_{\mathrm{NN}}} = 20-200\text{ GeV}$. By modifying the HIJING transport model to incorporate QCD-inspired partonic energy-loss formalisms for hot and cold nuclear media, we reduce reliance on the default HIJING energy-loss parameter and obtain a phenomenological framework for investigating its influence on multiplicity cumulants. This modification provides a more physically grounded framework for analyzing the dynamic properties of the dense matter produced in heavy-ion collisions. The main results of this study are as follows:
\begin{enumerate}
\item Within the present modified-HIJING framework, the separation between the considered energy-loss scenarios becomes more pronounced at the highest RHIC energies. Multiplicity fluctuations show the largest separation between the model scenarios, whereas net-charge and net-proton observables are comparatively less sensitive.

\item Within the implemented phase-allocation ansatz, charged-hadron multiplicity cumulants show a stronger non-linear response than the considered net-charge and net-proton observables. The ratio $S\sigma=C_3/C_2$ exhibits the most pronounced response.

\item Enlarging the idealized kinematic acceptance enhances the phase-allocation signal in multiplicity cumulants. In realistic measurements, the attainable lower $p_T$ threshold will limit this enhancement.
\end{enumerate}
These results emphasize the critical role of higher-order multiplicity cumulants and the careful selection of kinematic windows in experimental searches. The findings provide valuable baseline predictions for interpreting forthcoming data from heavy-ion programs at FAIR (CBM) and NICA (MPD)
dedicated to charting the QCD phase diagram and locating the critical endpoint.

\par\bigskip
\noindent\textbf{Funding.}
This work was supported by the Belarusian Republican Foundation for Fundamental Research, grant No. F24M-036.

\par\bigskip
\noindent\textbf{CONFLICT OF INTEREST}

The authors declare that they have no conflicts of interest.

\begin{myreferences}
\bibitem{Aoki06}
Y. Aoki, G. Endr\H{o}di, Z. Fodor, S. D. Katz, K. K. Szab\'o,
``The order of the quantum chromodynamics transition predicted by the
standard model of particle physics'',
Nature {\bf 443}, 675 (2006).

\bibitem{Sun18}
K.-J. Sun, L.-W. Chen, C. M. Ko, J. Pu, Z. Xu, ``Light nuclei production as a probe of the QCD phase diagram'', Phys. Lett. B {\bf 781}, 499 (2018).

\bibitem{Blume18}
C. Blume, ``Open Questions in the Understanding of Strangeness Production in HIC -- Experiment Perspective'', EPJ Web Conf. {\bf 171}, 03001 (2018).

\bibitem{Biswarup17}
B. Paul, ``Charmonium production in p--Pb collisions with ALICE at the LHC'', PoS(EPS-HEP2017) 182 (2017), arXiv:1710.09634 [nucl-ex].

\bibitem{Shuryak17}
E. Shuryak, ``Strongly coupled quark-gluon plasma in heavy ion collisions'', Rev. Mod. Phys. {\bf 89}, 035001 (2017).

\bibitem{Koch17}
P. Koch, B. M\"uller, J. Rafelski, ``From strangeness enhancement to quark-gluon plasma discovery'', Int. J. Mod. Phys. A {\bf 32}, 1730024 (2017).

\bibitem{Adare08}
A. Adare, S. Afanasiev, C. Aidala [et al.] (PHENIX Collaboration), ``Charged hadron multiplicity fluctuations in $\mathrm{Au}+\mathrm{Au}$ and $\mathrm{Cu}+\mathrm{Cu}$ collisions from $\sqrt{s_{NN}}=22.5$ to 200 GeV'', Phys. Rev. C {\bf 78}, 044902 (2008).

\bibitem{Gazdzicki08}
M. Ga\'zdzicki, ``Onset of deconfinement and critical point: NA49 and NA61/SHINE at the CERN SPS'', Eur. Phys. J. Spec. Top. {\bf 155}, 37 (2008).

\bibitem{Xu14}
N. Xu, ``An overview of STAR experimental results'', Nucl. Phys. A {\bf 931}, 1 (2014).

\bibitem{Luo15}
X. Luo, ``Fluctuations of conserved quantities in high energy nuclear collisions at RHIC'', J. Phys.: Conf. Ser. {\bf 599}, 012023 (2015).

\bibitem{ALICE21}
S. Acharya, D. Adamov\'a, A. Adler [et al.] (ALICE Collaboration), ``Charged-particle multiplicity fluctuations in Pb--Pb collisions at $\sqrt{s_{NN}}=2.76$ TeV'', Eur. Phys. J. C {\bf 81}, 1012 (2021).

\bibitem{CMS23}
CMS Collaboration, ``Net-charge fluctuations in PbPb collisions at $\sqrt{s_{NN}}=5.02$ TeV'', CMS Physics Analysis Summary CMS-PAS-HIN-22-005 (2023).

\bibitem{Kekelidze18}
V. D. Kekelidze, V. Kolesnikov, A. Sorin,
``BM@N and MPD experiments at NICA'', EPJ Web Conf. {\bf 171}, 12001 (2018).

\bibitem{FAIRNICA}
P. Senger
``Heavy-ion collisions at FAIR-NICA energies'',  Particles {\bf 4}, 2 (2021), 214-226.

\bibitem{Mohanty02}
A. K. Mohanty, P. Shukla, M. Gleiser, ``Nonperturbative effects in a rapidly expanding quark-gluon plasma'', Phys. Rev. C {\bf 65}, 034908 (2002).

\bibitem{Yang98}
C. B. Yang, X. Cai, ``Multiplicity difference correlators under first-order QGP phase transition'', J. Phys. G: Nucl. Part. Phys. {\bf 24}, 1957 (1998).

\bibitem{Hippert17}
M. Hippert, E. S. Fraga, ``Multiplicity fluctuations near the QCD critical point'', Phys. Rev. D {\bf 96}, 034011 (2017).

\bibitem{Bleicher99}
M. Bleicher, E. Zabrodin, C. Spieles [et al.], ``Relativistic hadron-hadron collisions in the ultra-relativistic quantum molecular dynamics model'', J. Phys. G: Nucl. Part. Phys. {\bf 25}, 1859 (1999).

\bibitem{Nara19}
Y. Nara, ``JAM: An event generator for high energy nuclear collisions'', EPJ Web Conf. {\bf 208}, 11004 (2019).

\bibitem{Armesto12}
N. Armesto, B. Cole, C. Gale [et al.], ``Comparison of jet quenching formalisms for a quark-gluon plasma `brick''', Phys. Rev. C {\bf 86}, 064904 (2012).

\bibitem{Gyulassy94}
M. Gyulassy, X.-N. Wang, ``HIJING 1.0: A Monte-Carlo Program for Parton and Particle Production in High Energy Hadronic and Nuclear Collisions'', Comput. Phys. Commun. {\bf 83}, 307 (1994).

\bibitem{Baier95}
R. Baier, Yu. L. Dokshitzer, S. Peign\'e, D. Schiff, ``Induced gluon radiation in a QCD medium'', Phys. Lett. B {\bf 345}, 277 (1995).

\bibitem{Baier97}
R. Baier, Yu. L. Dokshitzer, A. H. Mueller, S. Peign\'e, D. Schiff, ``Radiative energy loss of high energy quarks and gluons in a finite-volume quark-gluon plasma'', Nucl. Phys. B {\bf 483}, 291 (1997).

\bibitem{Mueller96}
A. H. Mueller, ``Energy loss and $p_T$-broadening of high energy partons in hot and cold QCD matter'', Nucl. Phys. A {\bf 610}, 459 (1996).

\bibitem{Stachel07}
J. Stachel, ``Hard probes to diagnose the QGP at the LHC'', Heavy Ion Physics Perspectives, conference talk, 12--14 September 2007.

\bibitem{Piegne08}
S. Peign\'e, ``Collisional energy loss of a fast parton in QGP'', AIP Conf. Proc. {\bf 1038}, 139 (2008).

\bibitem{Steffens06}
F. M. Steffens, ``The temperature dependence of running coupling'', Braz. J. Phys. {\bf 36}, 582 (2006).

\bibitem{Rusak20}
Y. A. Rusak, L. F. Babichev, ``Monte Carlo simulation of the 1st order hadron--QGP phase transition in heavy ion collisions using a parton model'', Proc. Natl. Acad. Sci. Belarus, Phys.-Math. Ser. {\bf 56}, 84 (2020).

\bibitem{Chomaz04}
Ph. Chomaz, M. Colonna, J. Randrup, ``Nuclear spinodal fragmentation'', Phys. Rep. {\bf 389}, 263 (2004).

\bibitem{Randrup09}
J. Randrup, ``Phase transition dynamics for baryon-dense matter'', Phys. Rev. C {\bf 79}, 054911 (2009).

\bibitem{Randrup10}
J. Randrup, ``Spinodal phase separation in relativistic nuclear collisions'', Phys. Rev. C {\bf 82}, 034902 (2010).

\bibitem{Landau76}
L. D. Landau, E. M. Lifshitz, \textit{Statistical Physics}, Part 1 (Nauka, Moscow, 1976).

\bibitem{Zhang22}
F. Si, Y. Zhang, ``Statistical uncertainty estimation of higher-order cumulants with finite efficiency and its application in heavy-ion collisions'', Phys. Rev. C {\bf 105}, 024907 (2022).

\bibitem{Chen17}
L.-Z. Chen, Y.-Y. Zhao, X. Pan, Z.-M. Li, Y.-F. Wu, ``High cumulants of conserved charges and their statistical uncertainties'', Chin. Phys. C {\bf 41}, 104103 (2017).

\bibitem{STAR25}
Y. Huang, ``Precision measurements of kinematic scan for fluctuations of (net-)proton multiplicity distributions in Au+Au collisions from RHIC-STAR'', EPJ Web Conf. {\bf 364}, 15009 (2026).

\bibitem{Luo2015CBWC}
X. Luo, ``Volume fluctuation and auto-correlation effects in the moment analysis
of net-proton multiplicity distributions in heavy-ion collisions at RHIC'',
Nucl. Part. Phys. {\bf 40}, 105104 (2013).

\bibitem{STAR2016Cumulants}
L. Adamczyk et al. (STAR Collaboration), ``Cumulants of net-proton, net-kaon, and net-charge multiplicity
distributions in Au+Au collisions at RHIC'',
Phys. Rev. C {\bf 93}, 021903(R) (2016).
\end{myreferences}

\end{document}